\documentclass[pdflatex,sn-mathphys-num]{sn-jnl}% Math and Physical Sciences Numbered Reference Style
\usepackage{graphicx}%
\usepackage{multirow}%
\usepackage{amsmath,amssymb,amsfonts}%
\usepackage{amsthm}%
\usepackage{mathrsfs}%
\usepackage[title]{appendix}%
\usepackage{xcolor}%
\usepackage{textcomp}%
\usepackage{manyfoot}%
\usepackage{booktabs}%
\usepackage{makecell}
\usepackage{algorithm}%
\usepackage{algorithmicx}%
\usepackage{algpseudocode}%
\usepackage{algorithm}      % 提供 algorithm 环境
\usepackage{algpseudocode}  % 提供 algorithmic 环境（注意：这里用 algpseudocode 而不是 algorithmic）
\usepackage{amsmath}        % 提供数学符号支持

\usepackage{listings}%
\theoremstyle{thmstyleone}%
\theoremstyle{thmstyletwo}%

\theoremstyle{thmstylethree}%

\begin{document}

\title[Article Title]{Adaptive traffic signal control optimization using a novel road partition and multi-channel state representation method}


\author[1]{\fnm{Maojiang} \sur{Deng}}\email{2312709279@qq.com}

\author*[1,2]{\fnm{Shoufeng} \sur{Lu}}\email{itslu@njtech.edu.cn}

\author[3]{\fnm{Jiazhao} \sur{Shi}}\email{jiazhaoshi1113@gmail.com}

\author[4]{\fnm{Wen} \sur{Zhang}}\email{1031159738@qq.com}

\affil*[1]{\orgdiv{School of Transportation Engineering}, \orgname{Nanjing Tech University}, \orgaddress{\street{30 Puzhu south road}, \city{Nanjing}, \postcode{211816}, \state{Jiangsu}, \country{China}}}

\affil[2]{\orgdiv{Institute of ITS and Traffic Management}, \orgname{Jiangsu Province Engineering Research Center of
Transportation Infrastructure Security Technology}, \orgaddress{\street{30 Puzhu south road}, \city{Nanjing}, \postcode{211816}, \state{Jiangsu}, \country{China}}}

\affil[3]{\orgdiv{Tandon School of Engineering}, \orgname{New York University}, \orgaddress{\street{6 MetroTech Center, Brooklyn}, \city{New York}, \postcode{11201}, \state{New York}, \country{United States}}}

\affil[4]{\orgdiv{Faculty of civil engineering and mechanics}, \orgname{Jiangsu University}, \orgaddress{\street{301 Xuefu road}, \city{Zhenjiang}, \postcode{212013}, \state{Jiangsu}, \country{China}}}


\abstract{This study proposes a novel adaptive traffic signal control method leveraging a Deep Q-Network (DQN) and Proximal Policy Optimization (PPO) to optimize signal timing by integrating variable cell length and multi-channel state representation. A road partition formula consisting of the sum of logarithmic and linear functions was proposed. The state variables are a vector composed of three channels: the number of vehicles, the average speed, and space occupancy. The set of available signal phases constitutes the action space, the selected phase is executed with a fixed green time. The reward function is formulated using the absolute values of key traffic state metrics—waiting time, speed, and fuel consumption. Each metric is normalized by a typical maximum value and assigned a weight that reflects its priority and optimization direction. The simulation results, using Sumo-TensorFlow-Python, demonstrate a cross-range transferability evaluation and show that the proposed variable cell length and multi-channel state representation method excels compared to fixed cell length in optimization performance.}

\keywords{traffic signal control, road partition, variable cell length, multi-channel state representation, deep Q network, proximal policy optimization}



\maketitle

\section{Introduction}\label{sec1}
\subsection{Background and significance}\label{subsec2}

Traffic congestion at urban intersections poses significant challenges to travel efficiency and exacerbates issues such as carbon emissions and energy consumption. Efficient regulation of traffic signal timing is a rapid and cost-effective method to mitigate intersection congestion. Advances in two areas have provided opportunities to improve signal control. The first is artificial intelligence, and the second is traffic flow sensor technology. 

Recent advances in artificial intelligence (AI) have set the stage for enhanced performance in signal timing optimization. Data-driven approaches, particularly those utilizing AI techniques such as reinforcement learning (RL), Deep Q-Networks (DQN), Proximal Policy Optimization (PPO) and the Actor-Critic method, have garnered increasing attention for their potential to improve traffic signal control. Haydari and Yilmaz (2022) provided a comprehensive overview of this field, examining the unique characteristics of state, action, and reward modeling approaches. Their analysis encompasses single-agent RL, multi-agent RL, and deep RL methods for traffic signal management. 

Advances in traffic flow sensor technology, particularly radar-video sensor, offer new opportunities for developing more complex and refined signal control models. Traditional traffic detectors typically measure aggregate metrics, such as flow rate, average speed, often overlooking the spatial position, and dynamic behaviors of vehicles. The Radar-video sensor can capture position-dependent traffic flow parameters such as flow rate, queue length, occupancy, and speed of individual vehicles. The current detection range of the radar-video sensor is from 150 to 500 meters, capable of capturing comprehensive traffic flow parameters for all vehicles. This technological advancement effectively addresses the limitations of traditional detectors, which typically suffer from limited coverage and insufficient parameter detection capabilities. So, the radar-video sensor has significantly improved and refined the data collection capabilities. 

With AI's advanced nonlinear mapping ability and the fine-grained data acquisition from radar-video detection systems, research on enhancing signal control optimization models becomes both scientifically meaningful and practically impactful in city traffic management.

\subsection{Literature review}
The construction of state representation, action definition, and reward formulation is pivotal in evaluating the performance of deep learning methods in traffic signal control. This section reviews the literature across these three dimensions, with a comparison of selected studies presented in Table 1 to highlight differences among various approaches. 

\begin{table}[h]
\caption{Comparison of RL approaches for signal control}\label{tab1}%
\begin{tabular}{@{}llllll@{}}
\toprule
RL approach in references & Road partition & State & \makecell[l]{Phase\\ sequence} & \makecell[l]{Number of\\ phases} & Reward\\
\midrule
\makecell[l]{Deep convolutional \\Q network\\ (Genders, Razavi 2016)} & Fixed length& \makecell[l]{Presence or \\absence, speed} & Variable & 4 & \makecell[l]{ Change in cumulative\\ vehicle delay}\\
DQN(Fan et al. 2023) & Not applicable & Queue length & Fixed & 4 & queue length difference\\
DQN(Jamil et al. 2021) & Not applicable & \makecell[l]{Waiting time,\\ queue length\\ and delay} & Fixed & 2 & \makecell[l]{Waiting time, queue\\ length, delay}\\
\makecell[l]{Value-Based DQN\\(Wang, Hwang 2018)} & Fixed length& \makecell[l]{Current phase,\\ green phase duration,\\ red phase duration,\\ left-turn proportion } & Variable & 8 & Delay\\
\makecell[l]{Dual Targeting Algorithm\\(Kodama et al. 2022)} & Not applicable & \makecell[l]{Current phase, \\occupancy and \\its variation,\\ the ratio \\between average \\speed and \\speed limit} & Variable & 4 & Cumulative delay\\
\makecell[l]{Teacher-Student DDQN\\(Liu et al. 2023)} & Not applicable & \makecell[l]{Number of cars, \\queue length, current phase} & Variable & 4 & Queue length\\
\makecell[l]{GBRT-RL\\(Savithramma 2023)} & Not applicable & Number of cars & Variable & Not provided & Delay\\
DQN(Guo et al. 2019) & Not applicable & \makecell[l]{Vehicles'\\ position and speed} & Variable & 8 & Queue length\\
DQN and PPO (this paper) & Variable length & \makecell[l]{Number of vehicles, \\average speed, \\and space occupancy} & Variable & 8 & \makecell[l]{Weighted sum \\of waiting time, \\speed and\\ fuel consumption}\\

\botrule
\end{tabular}

\end{table}

The evolution of state representation has transitioned from aggregated to disaggregated variables, and road partition from fixed to variable cells. Initially, aggregated variables like vehicle count, queue length, and delay were used. For instance, Li et al. (2016) employed queue length as a state variable, while Fan et al. (2023) considered both signal phase and queue length as states. Ma et al. (2022) advanced this by using deep neural networks to represent traffic conditions through sequences of images and employed an actor-critic model to determine traffic signal plans. Wang et al. (2024) enhanced the representation by integrating the CBAM model with the dueling double deep Q Network algorithm, improving the depiction of vehicle distribution and dynamics at intersections. Han et al. (2024) used a vector consisting of traffic counts on approaches and exits, the current phase as state, and applied the attention mechanism to approximate the value functions in the critic network.

With the advantages of no dimension compression and the strong feature representation ability of deep reinforcement learning, the state representation for traffic signal control shifts towards disaggregated state variables, marking a significant advancement. The precision level has greatly improved. The initial work by Genders and Razavi (2016) introduced Discrete Traffic State Encoding (DTSE), which used car position, speed, and signal phase as state vectors. Subsequently, Shabestary and Abdulhai (2022) partitioned roads into fixed cells to create matrices for car count and speed, which were then fused into image-like inputs. Shi et al. (2023) also used fixed cells, employing convolutional neural networks to encode vehicle presence and speed matrices. To ensure that the influence of cells, whether near or far, on signal timing is consistent, the variable cell partition approach is employed. Luo et al. (2020) proposed using the Fibonacci sequence for variable cell partitioning. This method adjusts cell sizes according to their distance from the intersection, thereby providing a more accurate representation of how different cells affect signal timing. Notably, the Fibonacci sequence is characterized by each term being the sum of the two preceding terms, which allows the length of other cells to be determined once the length of the first cell is established.

Key approaches in action definition include binary phases (Thorpe et al.(1996), Abdulhai et al.(2003), Wen et al(2007).), time adjustments within the green phase (Lu et al.(2008), and green phase duration (Toubhi et al.(2017)). The binary phase approach offers a choice between maintaining the current phase or switching to another phase at each time step.

Common indicators for defining rewards include delay, queue length, the number of waiting vehicles, and the frequency of stops. Reward functions are formulated based on these indicators to guide decision-making. For instance, Wan and Hwang (2018) introduced a dynamic discount factor within the iterative Bellman equation to mitigate biased estimations of the action-value function caused by inconsistent time-step intervals. Jamil et al. (2020) developed a composite reward architecture, leveraging the majority voting method and partial knowledge of nearby intersections to select optimal actions. Kodama et al. (2022) proposed a dual-targeting algorithm to enhance learning performance by reinforcing successful experiences in multi-agent environments. Liu et al. (2023) designed a single reward function for signal control to minimize hyper-parameters and model complexity within a teacher-student framework. Bouktif et al. (2023) emphasized that clearly defining both state and reward is crucial for achieving training stability and rapid convergence, and they proposed a double deep Q-network combined with a prioritized experience replay mechanism for the agent architecture. 

\subsection{Research gaps and contributions of this study}
The following research gaps are identified in the current study.

1.Current road partition methods fail to account for the detection range of sensors, resulting in a mismatch between the input parameters of trained AI models and the actual detection capabilities in practical applications. 

2. The AI model for signal control requires specialized training for each intersection, and its generalization ability and adaptability are poor.

This paper addresses research gaps by proposing a new road partition and multi-channel state representation method. 

1. A novel road partition formula, which combines logarithmic and linear functions, is proposed. Unlike traditional fixed-length road partition methods, the formula employs variable-length cells - shorter near stop lines for finer granularity and progressively longer with increasing distance. 
No matter how many meters the detection range of the detector is, this formula can discretize the road into the same number of cells. This significantly improves the AI model's generalizability and enables direct transfer of trained AI models to other intersections. 

2. A multi-channel state representation for traffic signal optimization is proposed, where the input state consists of three channels: the number of vehicles, average speed, and space occupancy within each cell. This multi-dimensional representation enables a more comprehensive understanding of the traffic system, leading to more effective signal timing strategies.

\section{Methodology}\label{sec2}

Inspired by the application of convolutional neural networks in image processing by Krizhevsky et al. (2017), the traffic state at an intersection is abstracted into a feature map represented as a three-dimensional tensor. This tensor is structured with dimensions corresponding to the number of lanes × the number of cells × the number of traffic state variables, which analogously align with the Height × Width × Channels dimensions used in image processing. Each element within this tensor stores quantifiable data describing the microscopic traffic conditions at a specific spatial location. Specifically, for each cell position along a lane, the three channels encode the following critical state variables: the number of vehicles, the average speed of those vehicles, and the space occupancy of the cell. This structured representation effectively transforms the spatial and temporal distribution of traffic flow into a format amenable to convolutional feature extraction, allowing the network to discern complex, localized patterns for optimized signal control decisions. 

The features are extracted from the three channels, respectively, through convolutional kernels, and the features from different channels are fused in the convolutional layers, generating more comprehensive features. The proposed methodology is illustrated in Fig. 1. The roads are divided into non-uniform cells. The inputs to the fully connected neural network consist of three matrices, representing the number of vehicles, the average speed, and the space occupancy, each of equal size. Space occupancy is the ratio of the sum of the lengths of all vehicles in a cell to the length of the cell. As the right turn lane is not signal controlled, it was not discretized. The action space is defined as eight discrete phases. The duration of the green time of each phase is fixed. The reward function is formulated using the absolute values of key traffic state metrics—waiting time, speed, and fuel consumption. Each metric is normalized by a typical maximum value and assigned a weight that reflects its priority and optimization direction. The following subsections detail the road partition method, the representation of the state space, the action space, and the reward function.

\begin{figure*}[!ht]
\centering{\includegraphics[width=\linewidth]{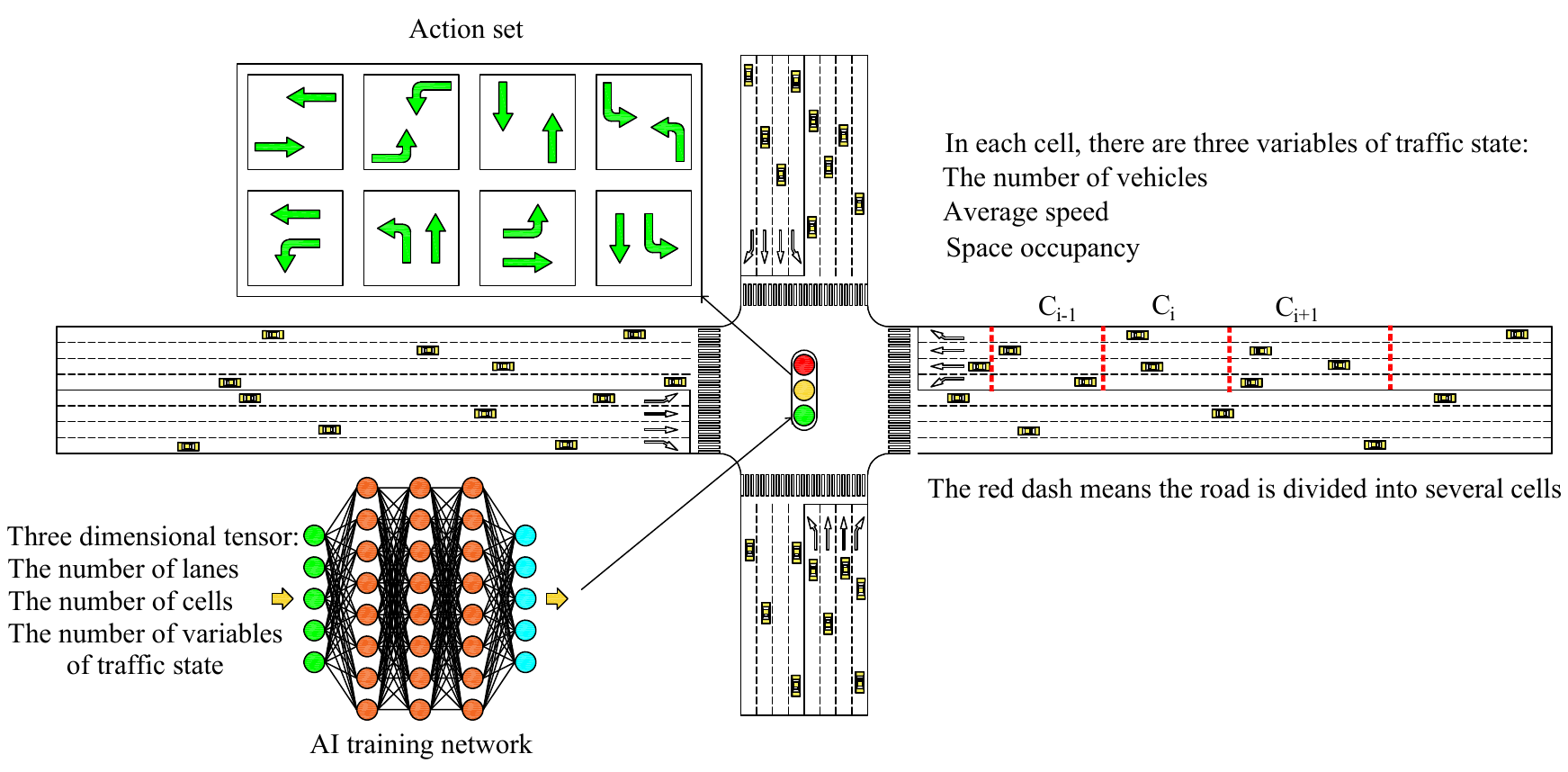}}
\caption{The structure of the proposed methodology}
\end{figure*}

\subsection{Road partition method}

A deep reinforcement learning-based adaptive signal control model consists of two stages: training and application. In the training stage, simulation software SUMO is typically used to construct the traffic network, vehicle inputs, sensors, and signal control, which offers considerable flexibility in the partitioning of the road and the placement of the traffic sensor. However, in the application stage, traffic data collection is restricted by the location of the installation and the performance of the traffic sensors. At urban intersections, sensors commonly used to optimize signal timing include loop, microwave, video, and radar-video sensors. A comparison of these types of sensor is shown in Table 2. The detection range of sensors varies greatly. To ensure consistency between the training stage and the application stage, the proposed road partition method must be able to adapt to the wide range of sensor measurements. 

\begin{table}[h]
\caption{Comparison of sensors}\label{tab1}%
\begin{tabular}{@{}llllll@{}}
\toprule

Sensor type  & Detection range  & Traffic parameters  \\
\midrule
Loop detectors & Fixed locations near the stop line
& Vehicle counts, speed, occupancy\\

Microwave detectors & Vertical range of around 160 meters  & \makecell[l]{Vehicle counts, speed, occupancy, vehicle type,\\ headway, queue length}\\

Video detectors & Usually between 5 and 20 meters & \makecell[l]{Vehicle counts, speed, occupancy,\\ vehicle type, vehicle color, presence, headway}\\

Radar-video detectors & Vertical range between 150 and 500 meters
& \makecell[l]{Vehicle counts, speed, occupancy, vehicle type,\\ vehicle color, presence, queue length}\\
\botrule
\end{tabular}

\end{table}

In this paper, a cell length function composed of a logarithmic function and a linear function is proposed. Upstream of the stop line, the cell length gradually increases. The reason for dividing cells in this way is that vehicles closer to the stop line have a greater impact on the signal timing scheme, whereas vehicles farther from the stop line have a smaller effect. Therefore, vehicles closer to the stop line need more precise treatment and detailed information on traffic condition, requiring shorter cell lengths. 

The length of each cell $f(x)$ is calculated using the proposed formula below.
\begin{equation}
f(x)  = a \ln(x+1)+bx
\label{eq1}\\
\end{equation} 
where $x$ is the cell index, $a$ and $b$ are the coefficients. 

The length of the first cell $l_1$ is predetermined. Usually, it is defined as the sum of the length of one car and the safe space headway, for example 7 meters.

\begin{equation}
f(1)  = a \ln2+b =l_{1}
\label{eq1}\\
\end{equation} 

Given the detection range $d$ of the sensor, the proposed function divides the detection range into $n$ cells. The sum of the lengths of the $n$ cells equals the detection range of the sensor.

\begin{equation}
d=\sum_{x=1}^{n}[a\ln(x+1)+bx]=a\sum_{x=1}^{n}\ln(x+1)+b\sum_{x=1}^{n}x
\label{eq1}\\
\end{equation} 

Under the condition that $l_1$ and $d$ are known, by solving equations (2) and (3) simultaneously, the values of $a$ and $b$ can be calculated.

\begin{equation}
a=\frac{d-l_{1}\sum_{x=1}^nx}{\sum_{x=1}^n\ln(x+1)-(\ln2)\sum_{x=1}^nx}
\label{eq1}\\
\end{equation} 

\begin{equation}
b=l_{1}-\frac{(\ln2)(d-l_{1}\sum_{x=1}^nx)}{\sum_{x=1}^n\ln(x+1)-(\ln2)\sum_{x=1}^nx}
\label{eq1}\\
\end{equation} 
where $\sum_{x=1}^n\ln(x+1)=\ln[(n+1)!]$, $\sum_{x=1}^nx=\frac{n(n+1)}{2}$.

Given the sensor detection range and the number of cells to be partitioned, the length of each cell can be calculated using formulas (1) to (5). Taking as an example, the division of an entrance road of the intersection into 10 cells.  The sensor detection range is 500 meters. The length of the first cell is set to 7 meters. The results of each cell length are shown in Table 3. The computed value for each cell length is rounded to the nearest whole. The length of the last cell is equal to the detection range minus the sum of the lengths of all previous cells.

\begin{table}[h]
\caption{Cell length }\label{tab1}%
\begin{tabular}{@{}lllllllllll@{}}
\toprule

Cell index  & 1  & 2 & 3 & 4 & 5 & 6 & 7 & 8 & 9 & 10 \\
\midrule
Computed value  & 7 & 15.6 & 24.87 & 34.49 & 44.33 & 54.34 & 64.46 & 74.67 & 84.95 & 95.28\\

Rounded value & 7 & 16 & 25 & 34 & 44 & 54 & 64 & 75 & 85 & 96\\
\botrule
\end{tabular}

\end{table}

The advantage of the proposed road partition method is that it can divide the road into a specified number of cells regardless of the detection range of the sensor. When the number of cells at different intersections is the same, a deep reinforcement learning model trained at one intersection can be transferred to another intersection, which improves the generalization of the AI-based signal control model.

Our proposed logarithmic-linear formula fundamentally differs from the Fibonacci sequence employed by Luo et al. (2020) in its flexibility and direct parameterization. Their method is a fixed pattern, which generates cell lengths according to a pre-defined Fibonacci sequence. While this creates increasing cell sizes with distance from the stop line, the entire sequence is determined once the first cell's length is fixed. Crucially, it cannot be directly adapted to a predetermined detection range or a specific number of cells without altering the fundamental Fibonacci pattern or resorting to scaling and truncation, which may not optimally utilize the available sensing range.

In contrast, our logarithmic-linear formula is adaptive partitioning, which dynamically calculates all cell lengths based on two direct inputs: the total detection range (d) and the desired number of cells (n). This ensures that the entire detection zone is partitioned exactly into n cells, with lengths increasing monotonically with distance. This provides superior flexibility and allows the partitioning scheme to be automatically tailored to different sensor configurations (varying d) and computational constraints (varying n), which is a significant practical advantage over fixed-pattern approaches.

\subsection{State space} 

The state space is represented by three identical size matrices, the number of vehicles, average speed, and space occupancy. To facilitate more efficient learning and convergence in the deep reinforcement learning model, the data from these matrices undergo maximum normalization, as outlined in Equation (6). 
\begin{equation}
x_{inorm}=\frac{x_i}{x_{maxhist}}
\label{eq1}\\
\end{equation}
In this equation, $x_i$ represents the raw data. $x_{inorm}$ is the normalized data, ranging from 0 to 1, and $x_{maxhist}$ denotes the maximum historical data value. The use of historical maximum data, rather than current data, is intended to differentiate traffic states under varying flow conditions and to ensure model convergence. 

\subsection{Action space} 
The action space consists of eight possible phases, as shown in Fig. 2: east-west straight, east-west left turn, north-south straight, north-south left turn, east straight and left turn, south straight and left turn, west straight and left turn, and north straight and left turn. The selected phase is executed with a fixed green time. The duration of the green time of a signal phase should be larger than the minimum green time. The minimum green time refers to the time required to ensure the safety of pedestrians crossing the street. According to Chinese design guideline, the average walking speed for pedestrians crossing the street is 1.1 meters per second. The minimum green time is obtained by dividing the length of the zebra crossing by the average walking speed. In the following case study, the selected phase is assigned a fixed green time of 15 seconds.

\begin{figure}[!ht]
\centering{\includegraphics{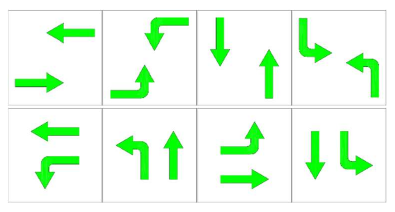}}
\caption{Eight actions}
\end{figure}

\subsection{Reward function}  
In a signalized traffic network, average waiting time, average speed, and average energy consumption are key indicators of congestion from the driver's perspective. The reward function is formulated using the absolute values of key traffic state metrics—waiting time, speed, and fuel consumption. Each metric is normalized by a typical maximum value and assigned a weight that reflects its priority and optimization direction: waiting time and fuel consumption carry negative weights to penalize their increase, while speed is assigned a positive weight to encourage its improvement. This design ensures that superior traffic conditions directly yield higher rewards, providing a consistent and intuitive learning signal for the agent throughout the training process. Normalization factors are applied to scale different metrics into comparable numerical ranges: waiting time divided by 200 seconds (typical maximum), speed divided by 15 m/s (typical maximum), fuel consumption divided by 50 mL (typical maximum). This ensures balanced contribution from each objective to the total reward. The reward function incorporates an explicit weighting scheme that reflects the relative importance of each objective. 

\begin{equation}
r(t) = -0.7 * waiting_{penalty} +  0.2 * speed_{reward} -0.1 * fuel_{penalty}
\end{equation}

where $waiting_{penalty} =  min(cur_{waiting} / 200.0, 1.0)$, $speed_{reward} = min(cur_{speed} / 20.0, 1.0)$, $fuel_{penalty} =  min(cur_{fuel} / 100.0, 1.0)$. 

The current waiting time $cur_{waiting}$, the current average speed $cur_{speed}$ and the energy consumption $cur_{fuel}$ are extracted from the Sumo software via the Traci interface. 

\section{Training algorithm}
Two training algorithms, the deep Q network (DQN) and the proximal policy optimization (PPO), are applied to compare their performance. Interactions between the DQN or PPO and the environment are stored in a replay buffer, recording states, actions, rewards, and the next states. After each training round, a subset of data is randomly sampled from this buffer to train the DQN or PPO, a process known as experience replay. This mechanism reduces training time and enhances efficiency, accelerating convergence. The size of the experience pool is set to 10000. Each time, 64 samples are randomly drawn from the experience pool.

\subsection{Deep Q network} 
A deep Q network (DQN) is used for adaptive control of the traffic signal. The TensorFlow framework, known for its robustness and flexibility, is utilized to build the DQN. The network consists mainly of fully connected dense layers, with dropout layers included for regularization to prevent overfitting. The ReLU (Rectified Linear Unit) activation function is used to speed up the training process. The DQN network architecture is shown in Table 4.
\begin{table}[h]
\caption{The DQN Network}\label{tab1}%
\begin{tabular}{@{}lllll@{}}
\toprule

 Layer  &  Role & Number of Neurons & \makecell[l]{The size of \\convolutional kernel} & Activation function\\
\midrule

Input layer & Input & $12\times10\times3=360$ 
&   &   \\
Hidden Layer 1 & Convolutional Layer & 32 & $3\times 3$  &  ReLU\\
Hidden Layer 2 & Pooling Layer &    &  $2\times 2$  &    \\ 
Hidden Layer 3 & Convolutional Layer &  64  & $3\times 3$ & ReLU \\
Hidden Layer 4 & Pooling Layer &    &  $2\times 2$  & \\
Hidden Layer 5 & Convolutional Layer &  128  &  $3\times 3$ & ReLU\\
Hidden Layer 6 & Pooling Layer &   &  $2\times 2$  \\
Hidden Layer 7 & Flatten Layer &    &    \\
Hidden Layer 8 & Fully Connected Layer & 128   &  &  ReLU  \\
Output Layer  & Output &  8  &  & Linear  \\
\botrule
\end{tabular}

\end{table}

For an intersection with 12 lanes for left turns and through traffic, each lane is divided into 10 cells. Each cell is described by matrices representing the number of vehicles, the average speed and the space occupancy, resulting in an input matrix of 1 row and 360 columns. The output matrix, with 1 row and 8 columns, holds the Q-values for the eight possible actions. The learning rate is set to 0.002 to balance stability and convergence speed during training.

\subsubsection{Action exploration}
The $\varepsilon$ greedy policy is used for action selection: with probability $\varepsilon$, a random action is chosen (exploration), and with probability $1-\varepsilon$, the action with the highest estimated reward is selected (exploitation). This strategy allows the agent to explore less frequently chosen actions while exploiting known rewarding actions. The 
$\varepsilon$ value decays linearly according to:

\begin{equation}
\varepsilon=1-\frac{e}{E}
\end{equation}
where $e$ is the current training episode number, $E$ denotes the total number of training episodes, and $\varepsilon$ represents the exploration rate for the current episode.

\subsubsection{Q-value update}
The Bellman equation is used to update Q values as follows: 
\begin{equation}
Q(s_{t},a_{t})=(1-\alpha)Q(s_{t},a_{t})+\alpha(r_{t}+\gamma \max_{a_{t+1}} Q(s_{t+1},a_{t+1}))
\end{equation}
where $Q(s_{t}, a_{t})$ is the Q-value for taking action $a_t$ in state $s_t$, and $\alpha$ is the learning rate, $r_t$ is the reward received after taking action $a_t$ in state $s_t$, $\gamma$ is the discount factor, $\max_{a_{t+1}}Q(s_{t+1}, a_{t+1})$ is the maximum Q-value over all possible actions that the agent can take from the next state $s_{t+1}$ after taking action $a_t$ in state $s_t$.

\subsubsection{The training process of DQN}

To enhance the generalization and adaptability of the model, the training process involves the stochastic generation of traffic flow patterns distributed randomly across 16 lanes. The traffic volume at the intersection ranges from 1600 to 2400 vehicles per hour with varied vehicle type distributions. Each training simulation lasts 10000 episodes, during which the same set of simulated traffic data is utilized in 25 episodes. Traffic flow parameters are updated every 25 episodes to improve the model's representation of dynamic traffic scenarios. Each chosen action is executed for a fixed seconds, which is between minimum green time and queue clearing green time. Queue length, delay time, waiting time, and average speed are measured via the Traci interface provided by SUMO, and data are collected every 60 seconds. The pseudo-code of DQN training is as follows.

\begin{algorithm}[H]
\caption{DQN Training}
\begin{algorithmic}[1]
\State Initialize DQN model;
\State Initialize the SUMO simulation environment;
\For{$epoch \gets 0$ \textbf{to} $epochs$}  
    \State Set action exploration rate, $\epsilon \gets 1.0$;
    \State Update traffic flow parameter;
    \For{$episode \gets 0$ \textbf{to} $episodes$}
        \State Update action exploration rate: $\epsilon \gets 1 - \frac{episode}{episodes}$;
        \For{\textbf{each} simulation step}
            \State Execute action $a$ in state $s$, observe reward $r$ and next state $s'$;
            \State Store tuple $(s, a, r, s')$ in DQN model's experience pool;
        \EndFor
        \State Sample a batch of experiences from the pool;
        \For{\textbf{each} tuple $(s, a, r, s')$ in batch}
            \State Compute $Q(s_{t}, a_{t})$ using DQN model;
            \State Update target: $Q_{\text{target}} \gets r_{t} + \gamma \max_{a_{t+1}} Q(s_{t+1}, a_{t+1})$;
            \State Update $Q(s_{t}, a_{t}) \gets (1 - \alpha) Q(s_{t}, a_{t}) + \alpha Q_{\text{target}}$;
            \State Update DQN model weights via backpropagation;
        \EndFor
    \EndFor
\EndFor
\State Save the DQN model;
\end{algorithmic}
\end{algorithm}
 
\subsection{Proximal policy optimization} 

Proximal policy optimization (PPO) employs a clipped surrogate objective function to ensure stable and reliable policy updates. This approach addresses the challenges associated with large policy updates by introducing a clipping mechanism that restricts the probability ratio between the new and old policies within a specified range. The clipping mechanism serves to prevent excessively large policy updates that could destabilize training. It ensures that the policy update remains within a trust region, thereby maintaining stability. The objective function in PPO is defined as follows.

\begin{equation}
L^{\text{CLIP}}(\theta) = \mathbb{E}_t \left[ \min \left( r_t(\theta) \hat{A}_t, \text{clip}(r_t(\theta), 1 - \epsilon, 1 + \epsilon) \hat{A}_t \right) \right]
\end{equation}
where $r_t(\theta)$ represents the probability ratio between the new policy 
$\pi_{\theta}$ and the old policy $\pi_{\theta_{\text{old}}}$ for action $a_t$ at state $s_t$. $\theta$ is a parameter of the policy network.

\begin{equation}
r_t(\theta) = \frac{\pi_{\theta}(a_t | s_t)}{\pi_{\theta_{\text{old}}}(a_t | s_t)}
\end{equation}

The term $\hat{A}_t$ denotes the estimated advantage value at the time step $t$, and $\epsilon$ is a small positive hyperparameter that defines the clipping range. 
The advantage function $\hat A_t$ is computed as:
\begin{equation}
\hat A_t = \delta_t + (\gamma\lambda)\delta_{t+1} + \cdots + (\gamma\lambda)^{T-t+1}\delta_{T-1}
\end{equation}
where $\delta_t = r_t + \gamma V(s_{t+1}) - V(s_t)$ is the temporal difference residual, $\gamma$ is the discount factor, and $\lambda$ controls bias-variance tradeoff.

The gradient of the expected return $J(\theta)$ and parameter update are computed as follows.
\begin{equation}
\nabla_\theta J(\theta) = \mathbb{E}_t \left[ \nabla_\theta \log \pi_\theta(a_t|s_t) \cdot \hat A_t \right]
\end{equation}
Where $\nabla_\theta J(\theta)$ is gradient of the expected return. $\pi_\theta$ is a policy parameterized by $\theta$. $Q^\pi$ is action-value function. 

The value of parameter $\theta$ is updated as follows.

\begin{equation}
\theta_{k+1} = \theta_k + \alpha \nabla_\theta J(\theta_k)
\end{equation}

PPO algorithm employs a dual-network architecture: one to predict the probability distribution of actions (policy function), referred to as the Actor, and the other to estimate the value of taking a specific action in a given state (value function), referred to as the Critic. The Critic network shares the same architecture as the actor, except that the number of neurons in the output layer is reduced to one. Softmax function is used to choose the action. The architecture of the actor network is shown in the Table 5.
\begin{table}[h]
\caption{The Actor and Critic Network}\label{tab1}%
\begin{tabular}{@{}lllll@{}}
\toprule

 Layer  &  Role & Number of Neurons & \makecell[l]{The size of \\convolutional kernel} & Activation function\\
\midrule

Input layer & Input & $12\times10\times3=360$
&  & \\
Hidden Layer 1 & Convolutional Layer & 32 & $2\times 2$ & ReLU\\
Hidden Layer 2 & Pooling Layer &    &  $2\times 2$  \\
Hidden Layer 3 & Convolutional Layer &  64  & $2\times 2$  & ReLU\\
Hidden Layer 4 & Pooling Layer &    &  $2\times 2$  \\
Hidden Layer 5 & Flatten Layer &    &   \\
Hidden Layer 6 & Fully Connected Layer &  128  &  & ReLU  \\
Hidden Layer 7 & Fully Connected Layer & 256   &  & ReLU  \\
Output Layer  & Output &  \makecell[l]{Actor 8 \\Critic 1}     &   & \makecell[l]{Softmax \\Linear} \\
\botrule
\end{tabular}

\end{table}

The actor network consists of two convolutional layers with 32, 64 convolutional kernels, respectively. Each kernel has a size of $2\times2$. A pooling layer follows each convolutional layer to reduce dimensionality, compress data and parameters, mitigate overfitting, and enhance the model's fault tolerance. The experience pool stores a total of 10,000 samples. Each time, 64 samples are randomly drawn from the experience pool. The discount factor is 0.99, the clipping rate is 0.2, and the learning rate of the neural network is set to 0.0003. Each chosen action is executed for some seconds, which is between the minimum green time and the queue clearance time. The pseudo-code of PPO training is as follows.
\begin{algorithm}[H]
\caption{PPO Training}
\begin{algorithmic}[1]
\State Initialize PPO model;
\State Initialize the SUMO simulation environment;
\For{$epoch \gets 0$ \textbf{to} $epochs$}
    \State Randomly generate the inputs of vehicles;
    \For{$episode \gets 0$ \textbf{to} $episodes$}
        \For{\textbf{each} simulation step}
            \If{Switch phase}
                \State Observe the environment state $s$;
                \State Take action $a$, observe next state $s'$ and reward $r$;
                \State Store $(s, a, r, s')$ in experience pool;
            \Else
                \State Keep the current phase;
            \EndIf
        \EndFor
        \State Sample a batch of experiences from the pool;
        \For{\textbf{each} tuple $(s, a, r, s')$ in batch}
            \State Compute state values $V(s)$ and $V(s')$;
            \State Calculate policy probability ratio $\rho(\theta)$;
            \State Clip $\rho(\theta)$ and compute losses $\mathcal{L}_{\text{actor}}$, $\mathcal{L}_{\text{critic}}$;
            \State Update actor and critic networks via gradient descent;
        \EndFor
    \EndFor
\EndFor
\State Save the PPO model;
\end{algorithmic}
\end{algorithm}

\section{Testbed} 
This study employs Sumo microscopic traffic simulation software to construct a simulation scenario for a single signal control intersection. A road network is shown in Fig. 3. 

\begin{figure}[!ht]
\centering
\includegraphics[width=\linewidth]{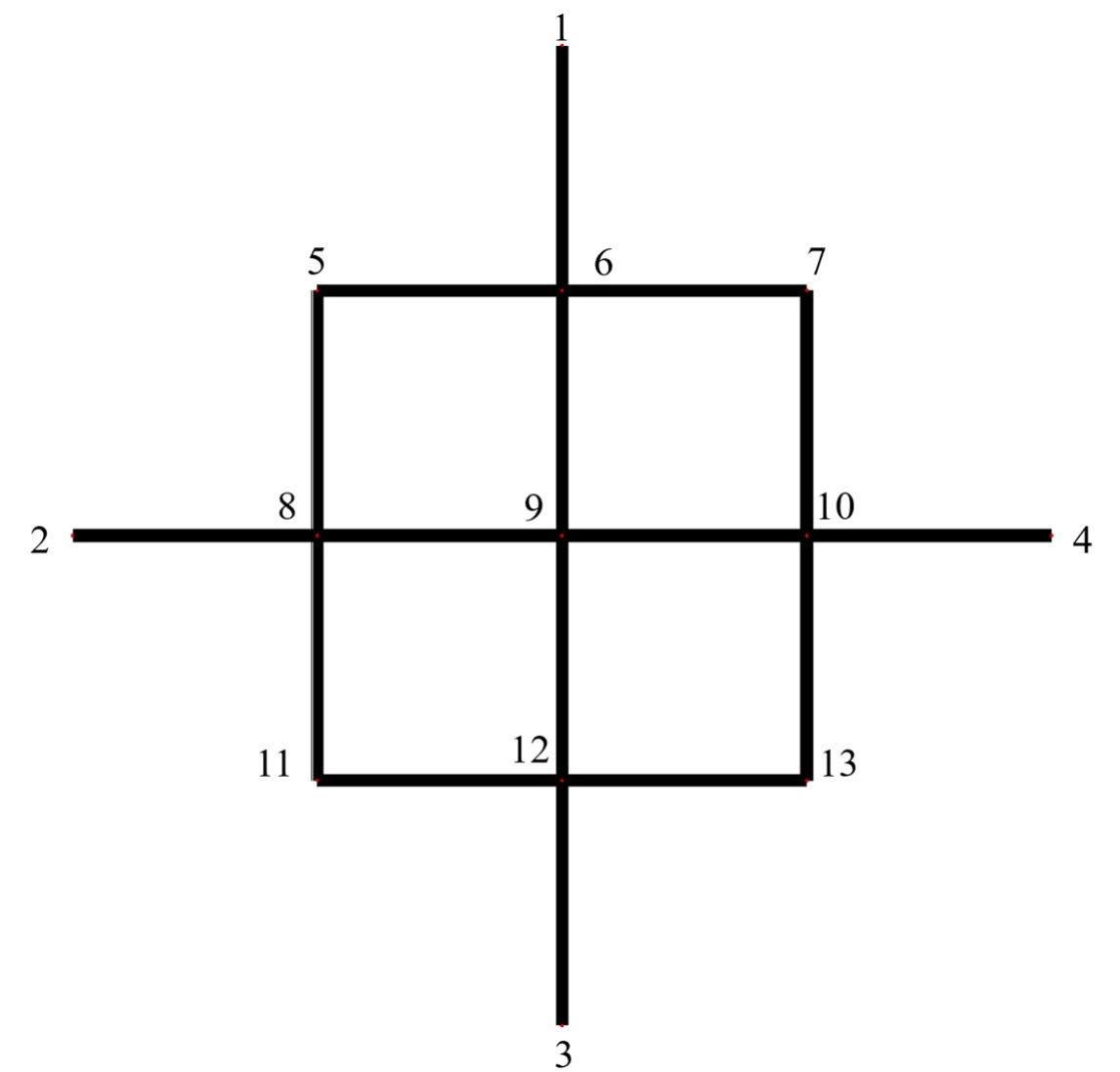} % 使用文本宽度
\caption{Road network}
\label{fig:roadnetwork}
\end{figure}

The numbers added to the SUMO road network represent the node IDs. Among them, nodes 1, 2, 3, and 4 denote traffic sources and sinks (origins and destinations), while nodes 5, 6, 7, 8, 10, 11, 12, and 13 are unsignalized intersections. Node 9 is the signalized intersection under study. The distance between the nodes is 500 meters. In the Python program, the adjacent nodes of each node are labeled in the form of a dictionary. 

To better align vehicle flow directions with real-world road networks, this section proposes a vehicle trajectory generation algorithm that randomly generates vehicle trajectories from sources to sinks. The vehicle generation is as follows. A random integer is uniformly sampled from the range [3, 6] and then multiplied by 400 to determine the total traffic flow for all four traffic sources (1, 2, 3, and 4) in Fig.3. This total flow is divided by 4 to calculate the average vehicle count per traffic source. For each source, the actual vehicle count is then randomly generated from a uniform distribution within the range [0.8 × average vehicle count, 1.2 × average vehicle count]. Trajectories corresponding to these vehicle counts are generated and the SUMO software loads these trajectories to simulate vehicles in the simulation. 

In the trajectory generation process, the road network is first topologically processed to label different node types and their adjacent nodes. A starting node is then randomly selected from the set of origin nodes to initiate a travel trajectory, while a corresponding end node is simultaneously initialized. Subsequently, adjacent nodes are randomly chosen based on the topological information starting from the origin node. If the selected node is a destination node, the trajectory is completed; otherwise, the process continues by selecting the next node until a destination node is reached. During the simulation, after generating random traffic flows, each vehicle is assigned a randomly generated travel trajectory according to the proposed vehicle trajectory generation algorithm. The pseudocode for the trajectory generation algorithm is as follows.

\begin{algorithm}[H]
\caption{Vehicle Trajectory Generation}
\begin{algorithmic}[1]
\State Initialize road network topology and node connections;
\For{each vehicle in total traffic flow}
    \State Randomly select a start node as origin;
    \State Add origin to trajectory list;
    \State Initialize $\mathit{previous\_node} \gets \text{None}$;
    \While{last node $\notin$ end nodes \textbf{or} trajectory length $< 2$}
        \State Filter possible next nodes based on last node and topology;
        \State Remove $\mathit{previous\_node}$ from candidate nodes;
        \If{no candidates available}
            \State \textbf{break};
        \Else
            \State Randomly select next node from candidates;
            \State Append selected node to trajectory list;
            \State Update $\mathit{previous\_node} \gets$ second-to-last node;
        \EndIf
    \EndWhile
    \State Generate complete trajectory;
\EndFor
\State Save all trajectories to file;
\end{algorithmic}
\end{algorithm}

The generated trajectories are vehicle inputs to the signalized intersection under study (node 9). Each approach at Node 9 includes one left-turn lane, two through lanes, and one right-turn lane, with a lane width of 3.5 meters. The speed limit on the cell is 72 km/h (20 m/s). The DQN and PPO are separately trained for 3000 episodes using TensorFlow 2.9.0 (GPU version) and Python 3.11 on a computer equipped with an AMD Ryzen 7940H processor and an NVIDIA GeForce RTX 4060 laptop GPU, with training taking approximately 2 to 3 hours.

For comparison with classical signal control methods, the DQN approach was benchmarked against fixed-time and actuated control strategies. The comparative results are presented in Fig. 4. The green time for a phase is computed as the ratio of the number of queuing vehicles (provided by the TraCI interface) to the average time headway. Experimental evaluations indicate that the proposed DQN model, incorporating variable-length cells, delivers performance that is considerably superior to that of conventional fixed-time and actuated control strategies.

\begin{figure}[!ht]
\centering{\includegraphics[width=12cm]{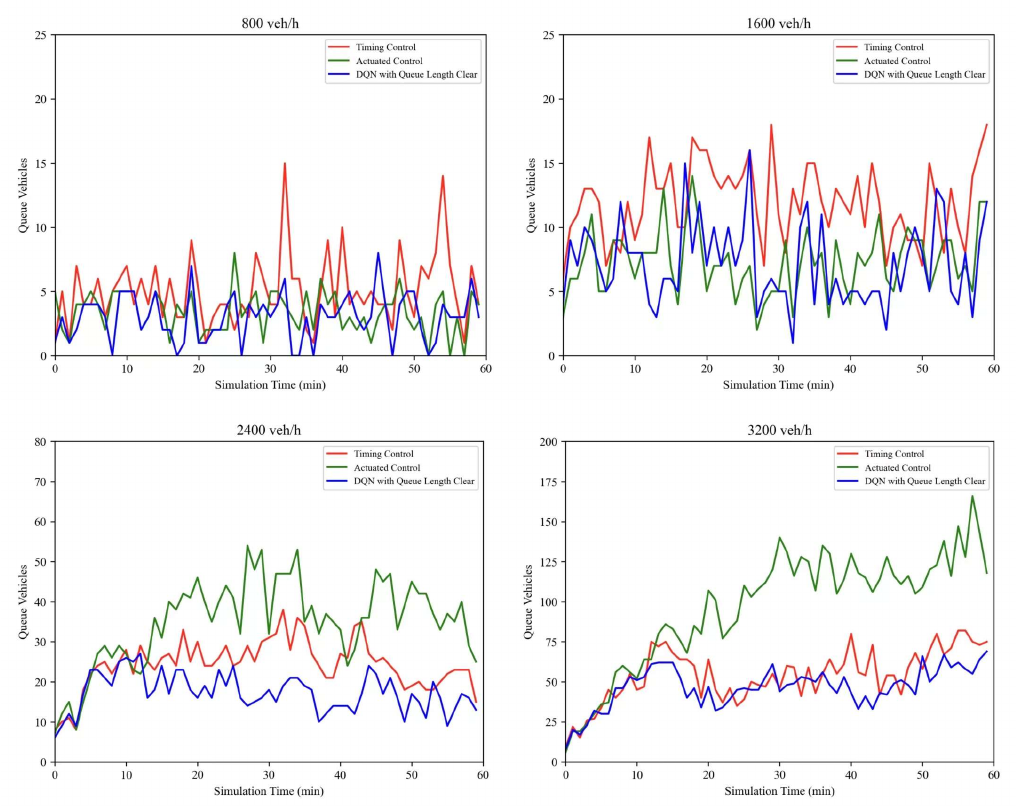}}
\caption{Comparison of queue lengths across four traffic scenarios}
\end{figure}

To evaluate the performance of the DQN and PPO model incorporating variable-length cells, simulation comparisons are performed in six traffic flow scenarios. The traffic volume is, respectively, 500, 1000, 1500, 2000, 2500 and 3000 veh/h. The traffic volume is defined as the sum of vehicles departing from the traffic sources consisting of nodes 1, 2, 3, 4. Run the trained DQN and PPO models for 2 hours in each scenario. The first half-hour is for warming up the simulation environment, and the data from the subsequent one and a half hours is used to evaluate performance. To minimize randomness and uncertainty in the experimental results, 10 simulation runs were performed for each scenario, and the average values were taken as the final results. During the training processes of DQN and PPO, each action (phase) may be selected multiple times, with a fixed green-light duration of 15 seconds for each occurrence. Upon convergence of the DQN and the PPO, the target network is used to evaluate performance. 

This section compares the performance of the proposed AI model—which integrates variable cell lengths and multi-channel state representation—with AI models using fixed cell length and AI models without cell-based discretization. The aim is to investigate the effectiveness of variable-length cell partitioning and multi-channel state description in improving traffic conditions. Fig. 5 and 6 present a comparative analysis of four signal timing optimization methods. 
In the figures, the PPO model incorporating variable cell length and multi-channel state representation is abbreviated as VCL PPO. The DQN model incorporating variable cell length and multi-channel state representation is abbreviated as VCL DQN. The PPO model with fixed cell length is denoted as FCL PPO, and the PPO model without cell discretization is simply referred to as PPO.

\begin{figure}[!ht]
\centering{\includegraphics{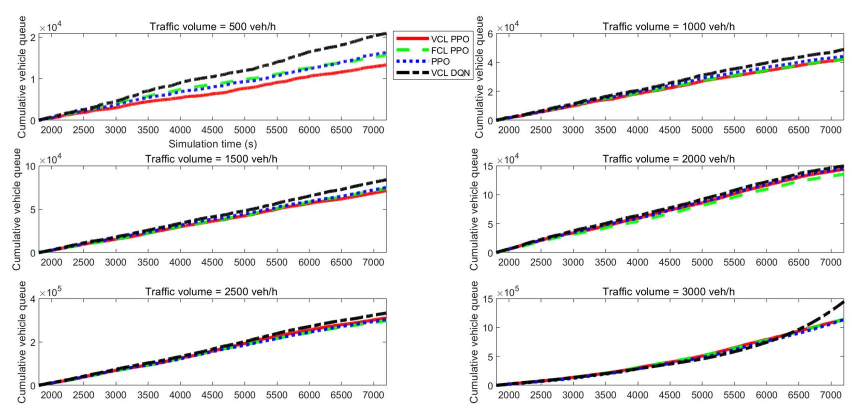}}
\caption{The comparison of cumulative vehicle queue}
\end{figure}

\begin{figure}[!ht]
\centering{\includegraphics{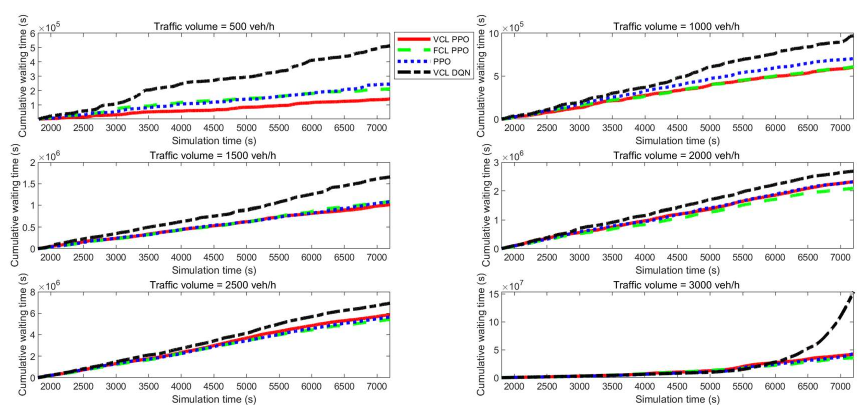}}
\caption{The comparison of cumulative waiting time}
\end{figure}

Fig. 5 uses cumulative queue length as the performance metric. Among the methods, VCL PPO achieves the best performance, whereas VCL DQN performs the worst. The other two algorithms exhibit intermediate performance. Similarly, in Fig. 6, which evaluates the cumulative waiting time, VCL PPO again outperforms the others, and VCL DQN remains the least effective. The performance gap between the four algorithms gradually narrows as the traffic volume increases. PPO consistently outperformed DQN in our experiments, primarily due to its superior stability and suitability for the signal control task. Unlike DQN, which estimates action values independently, PPO directly optimizes a parameterized policy, making it more robust to the noisy reward signals and sequential action constraints inherent in traffic phase optimization. Furthermore, PPO’s clipped objective prevents aggressive policy updates, ensuring steady learning progress even under non-stationary traffic conditions.

\section{Cross-Range Transferability Evaluation} 

\subsection{Experimental Design}
To rigorously evaluate the cross-range transferability of our proposed method, we designed a transfer test based on the following setup:

Source Scenario: A model was trained on an intersection with a sensor detection range of 300 meters, partitioned into 10 cells.

Target Scenario: The trained model was then applied directly to a different intersection with identical lane geometry but a detection range of 500 meters, also partitioned into 10 cells.

The core enabler of this transfer is our variable cell length (VCL) formulation. The formula automatically calculates the length of each cell based on the input parameters $d$ (total detection range) and $n$ (number of cells). This ensures that for a given $n$, the model's state representation is always a sequence of $n$ cells that non-uniformly partition the observable road space, regardless of the absolute physical scale. Consequently, a model trained with $n=10$ learns a control policy that operates on a relative spatial abstraction (e.g., reacting to vehicles in the 1st cell, the 2nd cell, etc.), which remains semantically consistent even when the physical distance covered by each cell changes from 300m to 500m. This decouples the learned policy from a fixed physical scale, enabling seamless transfer across different detection ranges.

\subsection{Experimental Results}
For a fixed flow rate of 1600 veh/h, which is the sum of flow rate from traffic origins 1,2,3,4 in Fig. 3, the learning curves using PPO for detection range 500m and 300m are shown in Fig. 7 and Fig. 8. The reward for detection 500m is higher than for detection 300m.
Both learning curves have good convergence.

\begin{figure}[!ht]
\centering{\includegraphics[width=12cm]{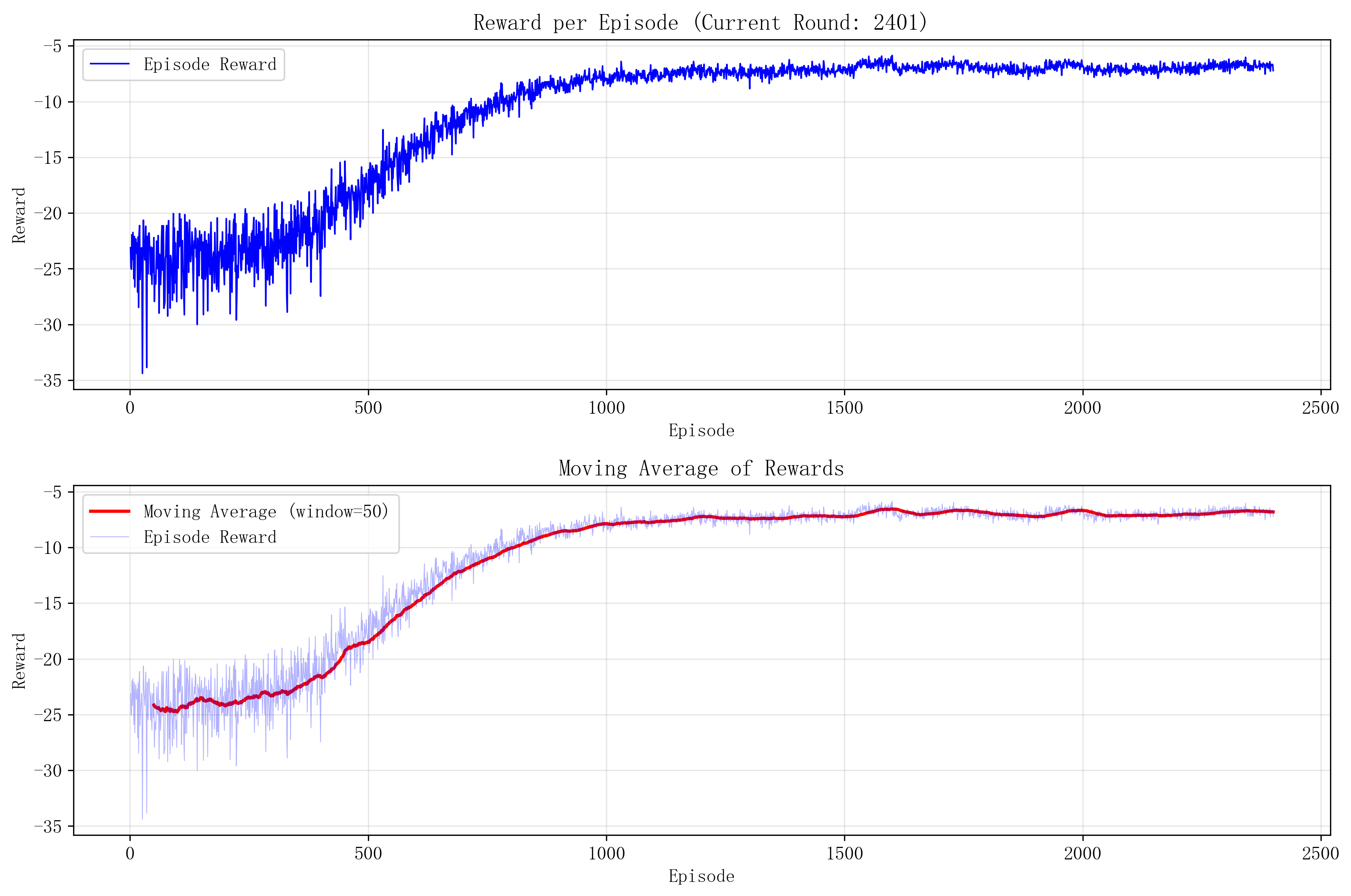}}
\caption{Learning curve using PPO for detection range 500m}
\end{figure}

\begin{figure}[!ht]
\centering{\includegraphics[width=12cm]{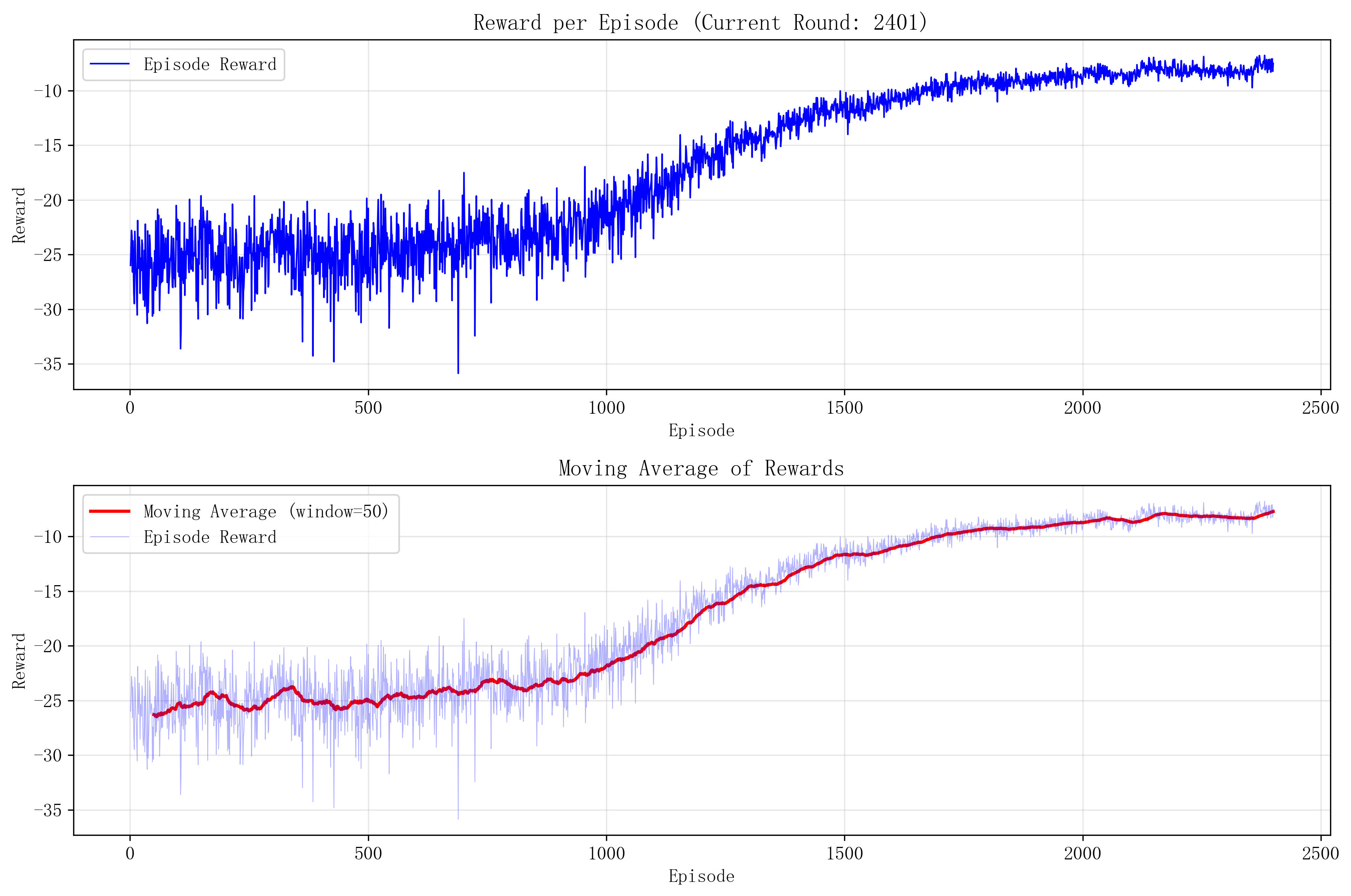}}
\caption{Learning curve using PPO for detection range 300m}
\end{figure}

A comparative analysis of model transferability is conducted by deploying a model trained with a 300-meter detection range to a road network designed for 500-meter detection. Performance metrics are evaluated across three control strategies: the model trained with 300-meter detection range, the model trained with 500-meter detection range, and a conventional fixed-time signal plan, to assess the adaptation capability of the models in unfamiliar operational environments. The simulation period is 1800 s. All data is about the intersection 9 in Fig. 3. The results are shown in Fig. 9. The results demonstrate that the VCL-PPO model trained on the 300m scenario, when directly transferred to the 500m scenario without any fine-tuning, maintained superior performance than fixed timing. 

\begin{figure}[!ht]
\centering{\includegraphics[width=12cm]{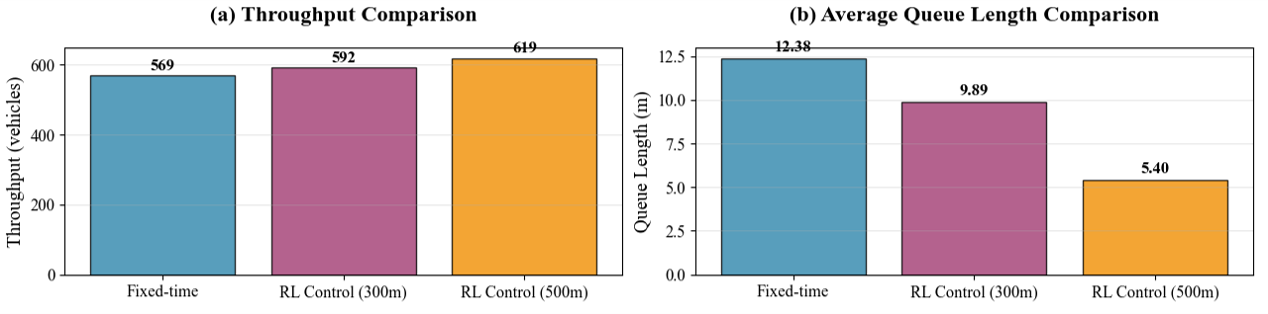}}
\caption{Performance comparison of cross-range transferability}
\end{figure}

For scenarios involving stochastic traffic flow variations, the flow rate was set to (3–5) × 400 veh/h for the initial 30 episodes, and subsequently varied between (2–7) × 400 veh/h. The learning curve for the PPO agent employing is presented in Fig. 10. Under these fluctuating traffic conditions, the reward progression exhibited significant oscillations, with a discernible and sustained improvement materializing only after approximately 1,000 episodes. This learning trajectory, characterized by its instability and slower convergence rate, indicates a substantially greater sample complexity compared to training under a fixed flow rate.

\begin{figure}[!ht]
\centering{\includegraphics[width=12cm]{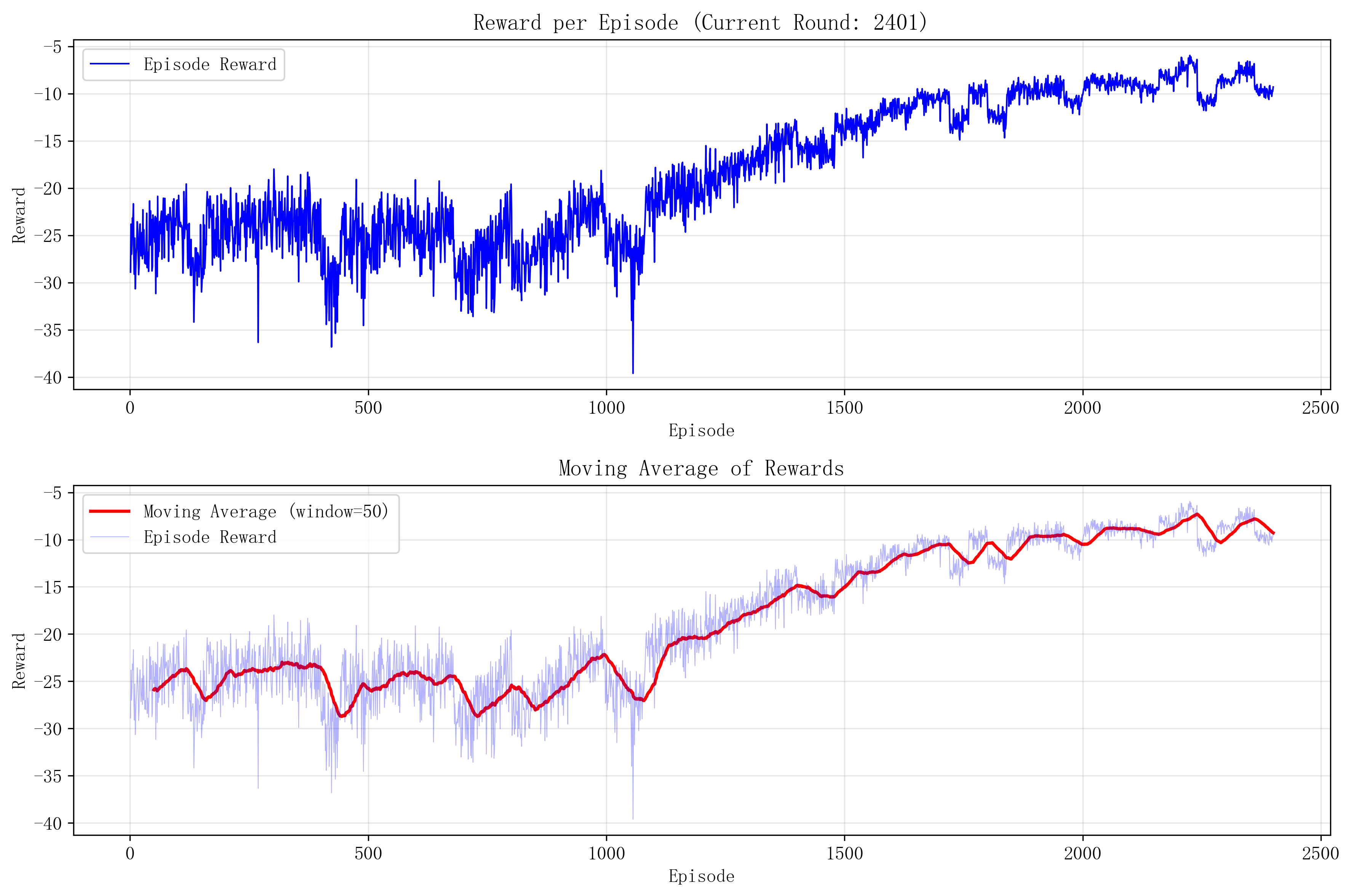}}
\caption{Learning curve using PPO for detection range 500m under stochastic traffic flow}
\end{figure}

\section{Conclusion and Discussion}

This paper proposes a signal timing optimization method that integrates variable cell length and multi-channel state representation. The design of this framework is inherently algorithm-agnostic, ensuring compatibility with a wide range of AI models. The proposed cell partitioning method, based on  the sum of logarithmic and linear functions, features a fixed number of lane divisions where each cell length adjusts automatically according to the detection range. This method ensures that the trained AI model can be universally applied across intersections with identical cell configurations and variable detection ranges. Under fluctuating traffic conditions, the reward progression exhibited significant oscillations, which needs more episode to converge compared to training under a fixed flow rate. In addition, to better align vehicle flow directions with real-world road networks, this paper
proposes a vehicle trajectory generation algorithm that randomly generates vehicle
trajectories from sources to sinks. In the presented case study of a single intersection, we conducted a comparative analysis focusing on two foundational deep reinforcement learning algorithms: DQN and PPO. The results indicate that the VCL-PPO model achieved the best performance in terms of both cumulative queue length and waiting time. This choice was made to first establish and clearly demonstrate the core contribution of the variable cell length integration within a controlled setting. The promising results from PPO suggest a robust foundation, and we explicitly acknowledge that exploring the integration of our framework with more sophisticated deep reinforcement learning methods (e.g., graph-based or attention-enhanced models) constitutes a critical and valuable direction for future research.

\vspace{10pt}\noindent\textbf{Authors’ contributions}\\
\noindent
Maojiang Deng: Literature Survey, Conceptualization, Methodology, Formulation, Coding(simulation).
Shoufeng Lu: Corresponding author, Literature Survey, Conceptualization, Methodology, Formulation, Coding(simulation), Writing, and Supervision.
Jiazhao Shi: Coding(simulation) and Visualization (Figure/Table Creation). 
Wen Zhang: Conceptualization.

\vspace{10pt}\noindent\textbf{Funding}\\
\noindent
The authors declare no funding support for this research.

\vspace{10pt}\noindent\textbf{Data Availability}\\
\noindent
Some or all data, models, or code that support the findings of this study are available from the corresponding author upon reasonable request.

\vspace{10pt}\noindent\textbf{Declarations}\\
\textbf{Competing interests}\\
\noindent
The authors declare that they have no known competing financial interests 
or personal relationships that could have appeared to influence the work 
reported in this paper.

\end{document}